\documentclass[10pt,letterpaper]{article}
\usepackage{pstricks}
\usepackage{graphicx}
\begin{document}

%%%%%%%%%%%%%%%%%% title page information %%%%%%%%%%%%%%%%%%
\title{Theory of Electro-optic Modulation via a Quantum Dot Coupled to a Nano-resonator}

\author{Arka Majumdar, Nicolas Manquest, Andrei Faraon and Jelena
Vu\v{c}kovi\'{c}\\
E. L. Ginzton Laboratory, Stanford University,\\
Stanford, CA, 94305\\
arkam@stanford.edu} %% email address is required
\maketitle



\begin{abstract}
In this paper, we analyze the performance of an electro-optic
modulator based on a single quantum dot strongly coupled to a
nano-resonator, where electrical control of the quantum dot
frequency is achieved via quantum confined Stark  effect. Using
realistic system parameters, we show that modulation speeds of a
few tens of GHz are achievable with this system,  while the energy
per switching operation can be as small as $0.5$ fJ. In addition,
we study the non-linear distortion, and the effect of pure quantum
dot dephasing on the performance of the modulator.
\end{abstract}

%\ocis{($250.4110$) Modulators; ($250.3750$) Optical logic devices; ($250.5590$) Quantum-well, -wire and -dot devices; ($350.4238$) Nanophotonics and photonic crystals.} % REPLACE WITH CORRECT OCIS CODES FOR YOUR ARTICLE

%%%%%%%%%%%%%%%%%%%%%%%% References %%%%%%%%%%%%%%%%%%%%%%%%%
%\bibliographystyle{unsrt}
%\bibliography{draft5_bibl}

%%%%%%%%%%%%%%%%%%%%%%%%%%  body  %%%%%%%%%%%%%%%%%%%%%%%%%%
\section{Introduction}
During the past decade there has been an extensive effort towards
building solid state optical devices with micron-scale footprint.
The main drive behind this research is the development of optical
networks for chip to chip interconnects (\cite
{article_lipson_ring}, \cite{2004.paniccia_intel_nature},
\cite{2008.Vlasov_IBM_nat_phot}) and optical quantum information
processing (\cite {2007.JOBrien.Science},
\cite{2008.JOBrien.Science}, \cite{2008.Englund.transfer},
\cite{2008.Andrei_DIT_WG}). In these networks, modulators play an
essential role in sending and routing information. Most of the
micron-scale modulators fabricated so far are based on either
classical effects or collective quantum effects. However, a few
papers recently published (\cite {article:Andrei09},
\cite{2009.Englund.Modulator.PIN}) show that a single quantum
emitter can be deterministically used to control the transmission
of a beam of light.

The reduction in operating power is one of the most important
reasons for miniaturizing optical components down to scales where
the active component is represented by a single quantum emitter.
Currently, one of the main bottlenecks in developing high speed
electronic devices is the large loss in the metal interconnect at
high frequency (\cite {article:miller08}). The development of
light switches operating at levels of just a few quanta of energy
may be a solution. At the same time, these devices represent a new
tool in the rapidly developing toolbox of quantum technologies.

Our recent demonstration of electro-optic switching at the quantum
limit is based on a single InAs quantum dots (QDs) embedded in a
GaAs photonic crystal resonator. The system operates in the strong
coupling regime of cavity quantum electrodynamics (CQED). The
control of the quantum dot is achieved by changing the quantum dot
frequency via the quantum confined Stark effect
\cite{article:Andrei09}. In this paper we theoretically analyze
the performance of this type of electro-optic modulator. We study
its maximum achievable modulation speed, linearity, step response
and power requirement. Lastly, we investigate how pure QD
dephasing affects its performance.

\section{Modulation Method and Analysis}
The Master equation describing the dynamics of a single QD
(lowering operator $\sigma=| g\rangle \langle e|$; where
$|g\rangle$ and $|e\rangle$ are the ground and the excited states
of the QD) coupled to a single cavity mode (with annihilation
operator $a$) is given by $(\hbar=1)$ \cite{book:gar05}
\begin{equation}
\label{Maseq} \frac{d\rho}{dt}=-i[H,\rho]+ \kappa
\mathcal{L}[a]+\gamma \mathcal{L}[\sigma]+
\frac{\gamma_d}{2}(\sigma_z\rho\sigma_z-\rho)
\end{equation}
where $\rho$ is the density matrix of the coupled cavity/QD
system; $\gamma$ and $\kappa$ are the QD spontaneous emission rate
and the cavity population decay rate respectively; $\gamma_d$ is
the pure dephasing rate of the QD;
$\sigma_z=[\sigma^\dag,\sigma]$. $\mathcal{L}[D]$ is the Lindbald
operator corresponding to a collapse operator $D$. This is used to
model the incoherent decays and is given by:
\begin{equation}
\mathcal{L}[D]= 2D\rho D^\dag-D^\dag D \rho-\rho D^\dag D
\end{equation}
$H$ is the Hamiltonian of the system without considering the
losses and is given by (in rotating wave approximation, where the
frame is rotating with the driving laser frequency)
\begin{equation}
H=\Delta\omega_c a^\dag a +\Delta\omega_a \sigma^\dag
\sigma+ig(a^\dag \sigma -a\sigma^\dag)+\Omega(a+a^\dag)
\end{equation}
where $\Delta\omega_c = \omega_c-\omega_l$ and $\Delta\omega_a =
\omega_a-\omega_l$ are respectively the cavity and dot detuning
from the driving laser; $\omega_c$, $\omega_a$ and $\omega_l$ are
the cavity resonance, the QD resonance and the driving laser
frequency respectively; $\Omega$ is the Rabi frequency of the
driving laser and is given by $\vec{\mu}\cdot \vec{E}$, where
$\vec{\mu}$ is the dipole moment of the transition driven by the
laser and $\vec{E}$ is the electric field of the laser inside the
cavity at the position of the QD; and $g$ is the interaction
strength of the QD with the cavity.

\begin{figure}
\centering
    %\psgrid
    \rput(-.2,4.6){\large (a)}
    \rput(6,4.6){\large (b)}
    \includegraphics[scale=0.4]{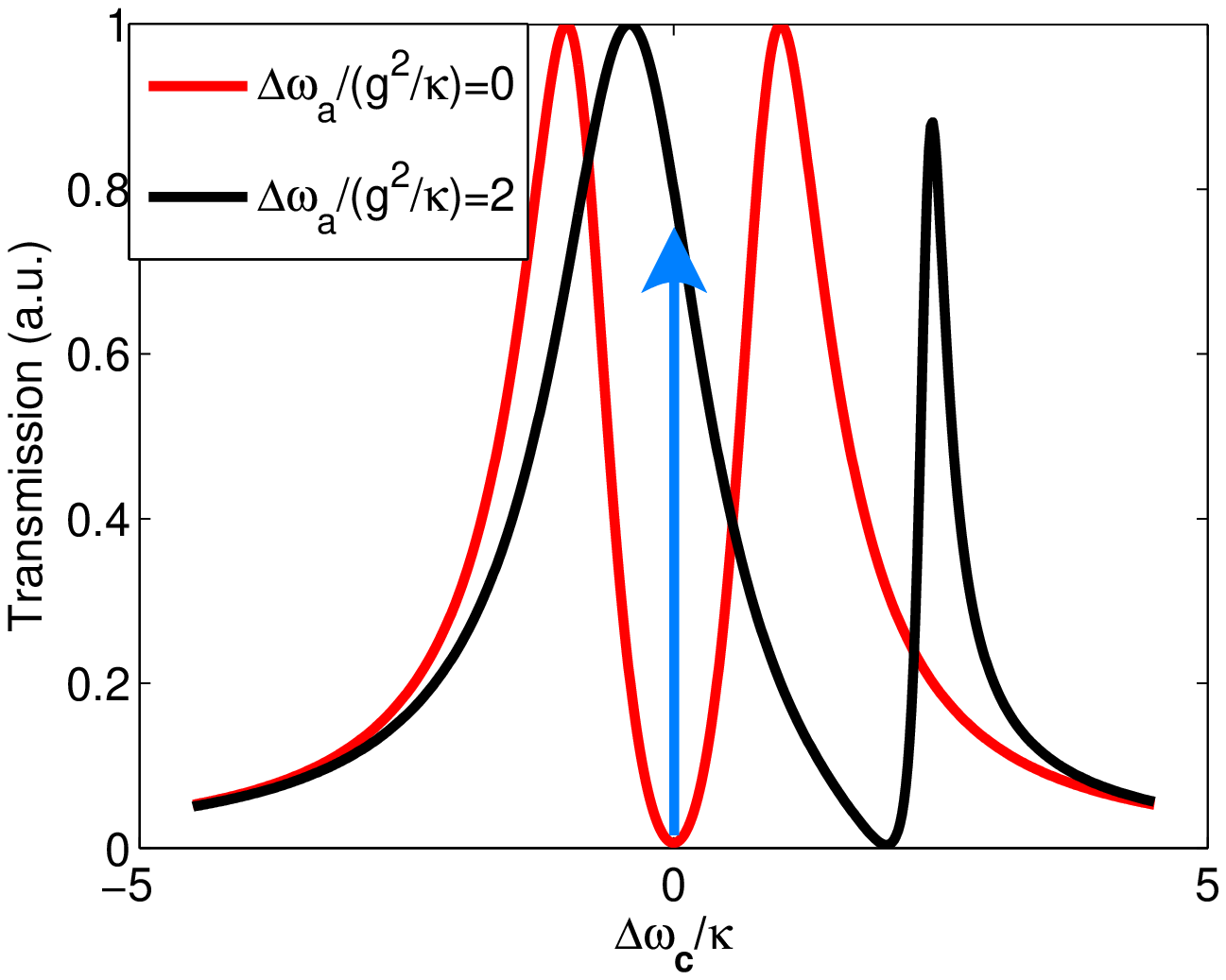}\includegraphics[scale=0.38]{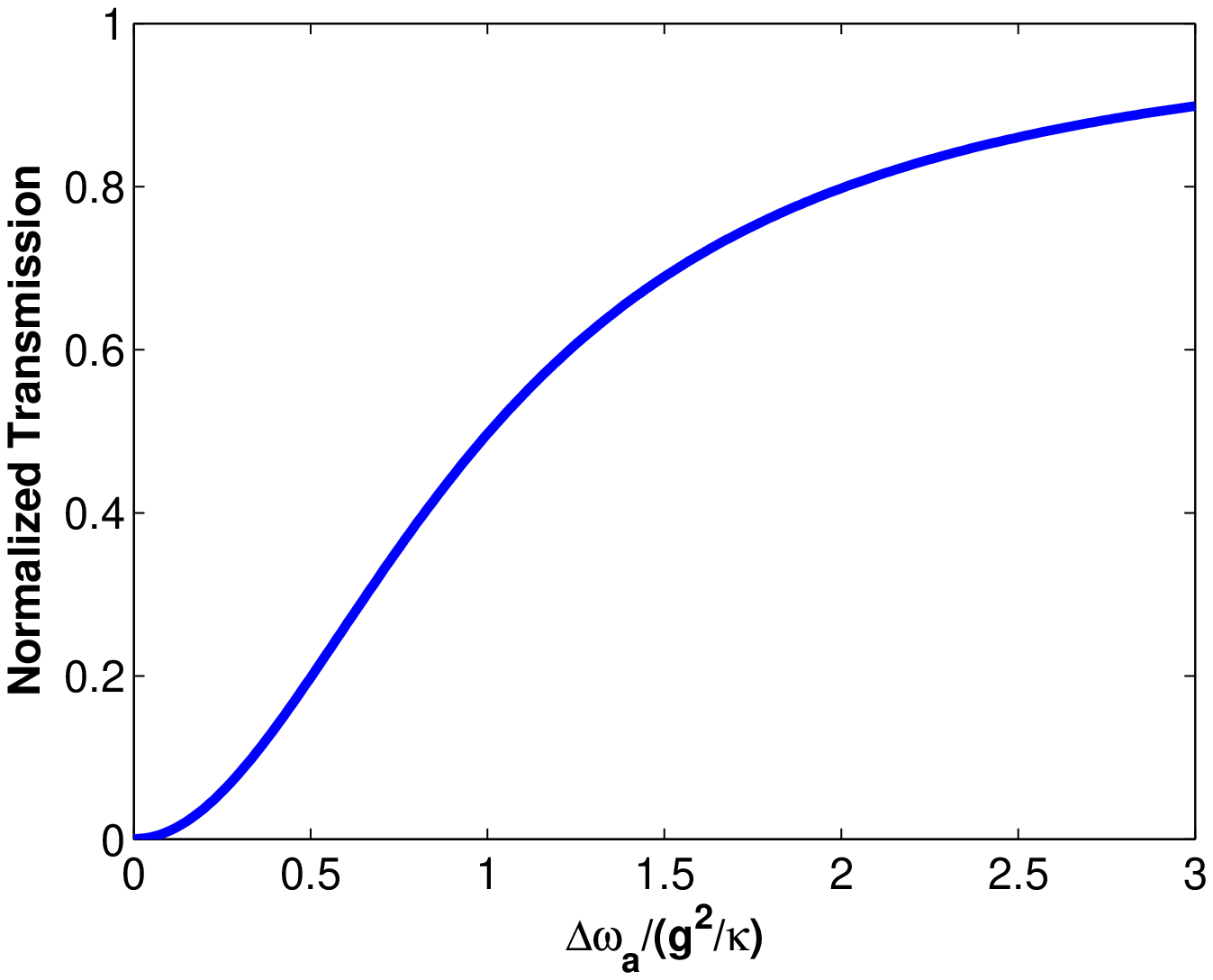}
\caption{(a) Transmission spectra of coupled cavity-QD system for
two different detunings. The blue arrow shows the wavelength of
the laser whose transmission is being modified. (b) Normalized
steady state transmission with different QD detunings. The laser
is resonant with the cavity. We used the following parameters:
$g/2\pi=\kappa/2\pi=20$ GHz; $\gamma=\kappa/80$; the pure
dephasing rate is assumed to be zero for this analysis
($\gamma_d/2\pi=0$). } \label{fig:scheme}
\end{figure}
Using the identity $\langle
\dot{\hat{O}}\rangle=Tr(\hat{O}\dot{\rho})$ for an operator
$\hat{O}$ and a density matrix $\rho$ in eq. \ref{Maseq}, we find
the following Maxwell-Bloch equations (MBEs) for the coupled
cavity/QD system (assuming the laser is resonant with the cavity
frequency, i.e., $\Delta\omega_c = 0$):
\begin{equation}
\label{MBE1}
  \frac{d\langle X\rangle}{dt} = A\langle X\rangle + i\Omega B
\end{equation}
\begin{equation}
\label{MBE2}
  \frac{d\langle Y\rangle}{dt} = C\langle Y\rangle + i\Omega D\langle X\rangle
\end{equation}
where
\begin{eqnarray}
  X &=& \left[%
\begin{array}{cccc}
  a & \sigma & a^\dag & \sigma^\dag \\
\end{array}%
\right]^T\\
 Y &=& \left[%
\begin{array}{cccc}
  a^\dag a & \sigma^\dag\sigma & a^\dag\sigma & a\sigma^\dag \\
\end{array}%
\right]^T\\
A &=& \left[\begin{array}{cccc}
        -\kappa & g  & 0 & 0\\
        -g & \Gamma & 0 & 0\\
        0 & 0 & -\kappa & g \\
        0 & 0 & -g & \Gamma^* \\
      \end{array}\right]\\
      B &=& \left[%
\begin{array}{cccc}
  -1 & 0 & 1 & 0 \\
\end{array}%
\right]^T\\
C &=& \left[\begin{array}{cccc}
        -2\kappa & 0  & g & g\\
        0 & -2\gamma & -g & -g\\
        -g & g & \Gamma-\kappa  & 0 \\
        -g & g & 0 & \Gamma^*-\kappa \\
      \end{array}\right]\\
D &=& \left[\begin{array}{cccc}
        1 & 0  & -1 & 0\\
        0 & 0 & 0 & 0\\
        0 & 1 & 0  & 0 \\
        0 & 0 & 0  & -1
      \end{array}\right]
\end{eqnarray}
where $\Gamma=-(\gamma+\gamma_d+i\Delta\omega_a)$. In deriving the
MBEs, we assume that under low excitation, the system stays mostly
in the lowest manifolds (single quantum of energy) and hence
$\langle a\sigma_z\rangle\approx-\langle a\rangle$ and $\langle
a^\dag a\sigma_z\rangle\approx-\langle a^\dag a\rangle$
\cite{article:edo_dit}.

When operating as a modulator, the cavity/QD system modulates the
transmission of the laser driving the cavity (laser amplitude $E$
and frequency $\omega_l$). The system is modulated using an
electrical signal which changes the QD resonance frequency via
quantum confined Stark effect. In our analysis, we assume that the
optical signal is always resonant with the bare cavity $(\omega_l
= \omega_c)$ and only the QD resonance frequency changes.

The system operates in strong coupling regime when
$g>>\kappa,\gamma$. In this regime, when the quantum dot and the
cavity are resonant, the optical system has a split resonance (in
contrast to a single Lorentzian), as shown in Fig.
\ref{fig:scheme}a. The detuning of the quantum dot due to the
application of an electric field causes dramatic changes in the
transmission spectrum of the optical system. This directly affects
the transmitted intensity of a laser tuned at the bare cavity
resonance (shown by an arrow in Fig. \ref{fig:scheme}a). Fig.
\ref{fig:scheme}b shows the steady-state transmission (normalized
by the transmission through an empty cavity) for different values
of QD-cavity detuning. This has been derived by solving the MBEs
(Eqs. \ref{MBE1}, \ref{MBE2}) at steady-state (in absence of pure
dephasing, i.e., $\gamma_d/2\pi =0$), which gives that the ratio
of the maximum to the minimum transmission through the cavity is
$\left(1+g^2/\kappa\gamma\right)^2$.

The performance of the modulator is analyzed by numerically
solving the MBEs, and thus deriving the time evolution of the
system. An alternative method involves numerically solving the
exact Master equation (eq. \ref{Maseq}). Both methods give us
exactly the same solutions and we report only the result obtained
from the MBEs, as this method is much faster and computationally
less demanding. To relate our theory to current state of the art
technology where values of $\kappa/2\pi = g/2\pi \approx 20$ GHz
can be easily achieved \cite{article:eng07}, we mainly analyze the
system performance for both $\kappa/2\pi$ and $g/2\pi$ ranging
between $10$ to $40$ GHz.

\section{Frequency Response}
The foremost criteria of a good electro-optic modulator is its
speed of operation. The modulation speed was analyzed by applying
a sinusoidal change in the QD resonance frequency:
$\Delta\omega_a(t) = \frac{1}{2}\Delta\omega_0(1-cos(\omega_e
t))$; where $\Delta\omega_0$ is the maximum detuning of the QD
resonance (this is proportional to the amplitude of the electrical
signal applied to tune the QD) and $\omega_e$ is the frequency of
the modulating electrical signal and hence also the frequency of
the change in the QD resonance. The system performance is
determined by analyzing the change in the cavity output (i.e.
$\kappa \langle a(t)^\dag a(t)\rangle$) as a function of
$\omega_e$ . The on/off ratio is defined as the ratio of the
maximum to minimum cavity output during sinusoidal driving.
 The Rabi frequency $\Omega$ of the driving laser
is chosen such that the QD is not saturated. The effect of free
carriers generation in semiconductor surrounding the dot by two
photon absorption of the driving laser is not included in our
analysis as at low $\Omega$ (that is, at low intensity of the
driving laser) this effect is very small. The pure dephasing rate
$\gamma_d/2\pi$ of the QD is assumed to be $0.1$ GHz.

\begin{figure}
\centering
    \rput(-.2,4.8){\large (a)}
    \rput(6,4.8){\large (b)}
     \includegraphics[scale=0.4]{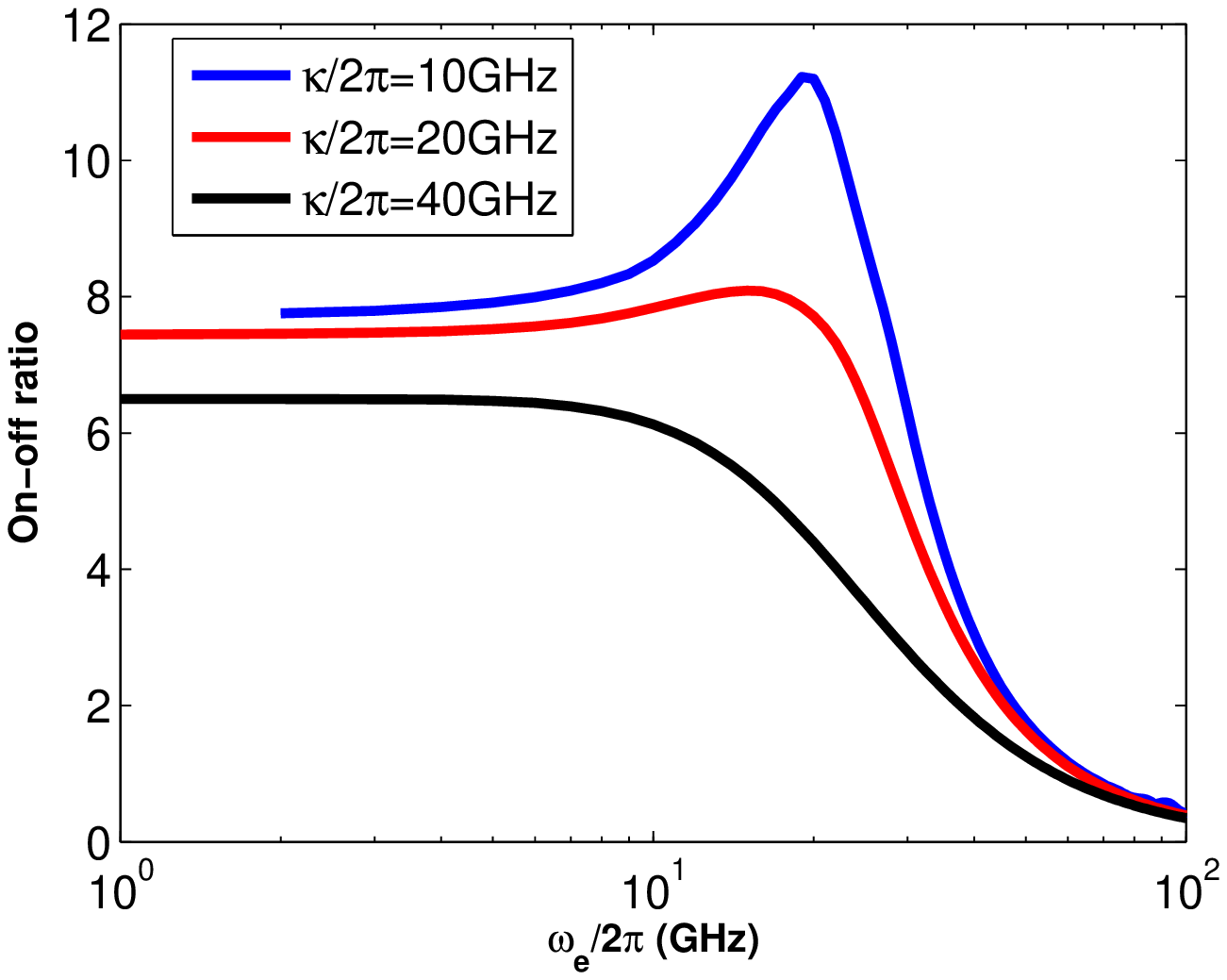}\includegraphics[scale=0.4]{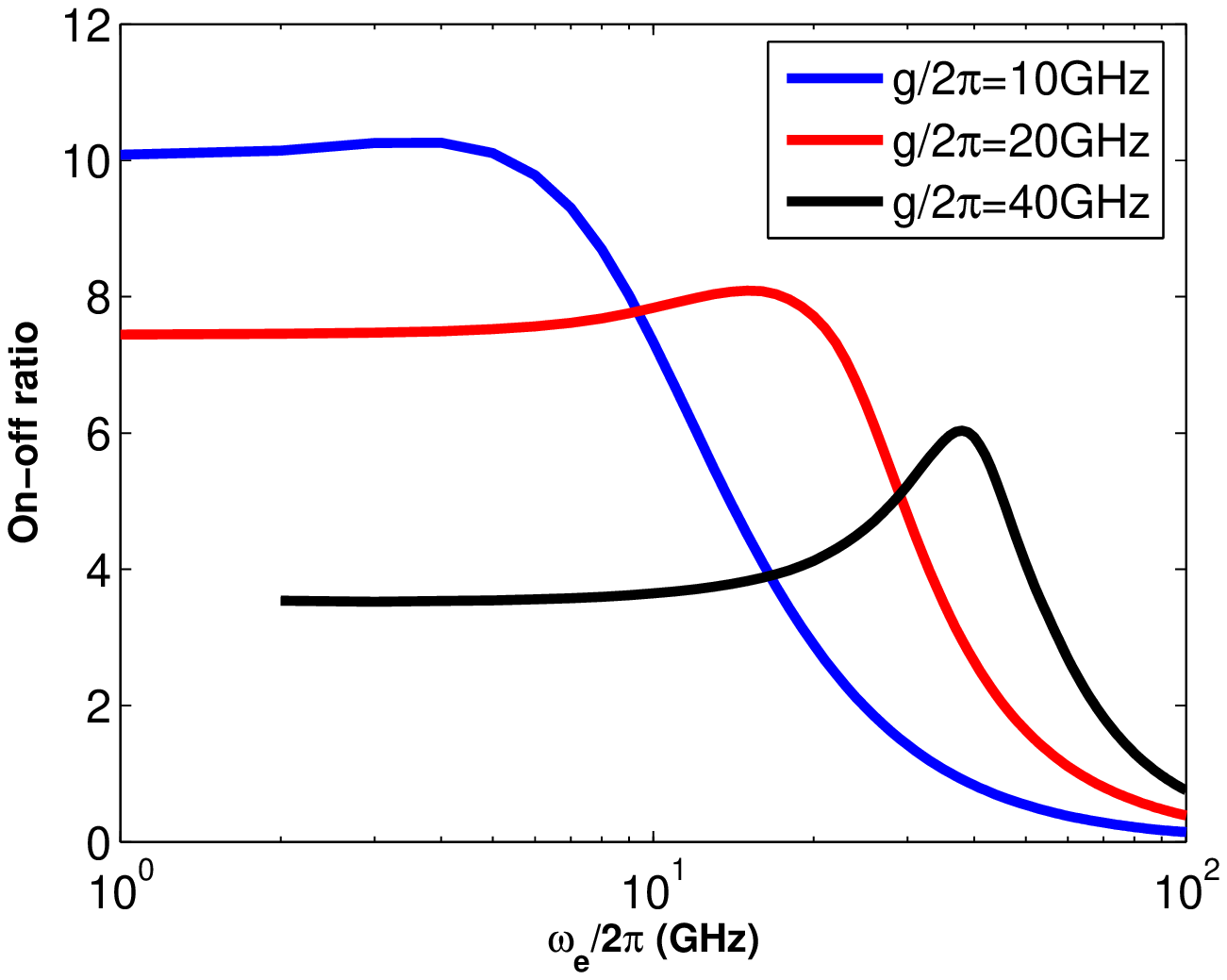}
\caption{The frequency response of the switch for different (a)
$\kappa$. $(g/2\pi)$ is kept constant at $20$ GHz and (b) g.
$(\kappa/2\pi)$ is kept constant at $20$ GHz. For both simulations
$\Omega = 1$ GHz; $\Delta\omega_0/2\pi = 10$ GHz; $\gamma_d/2\pi =
\gamma/2\pi = 0.1$ GHz.} \label{fig:freq_response_mb}
\end{figure}
\begin{figure}
\centering
    %\includegraphics[scale=1]{fig:freq_response.pdf}
    %\psgrid
    \rput(-.25,5){\large (a)}
    %\rput(6.4,4.7){\large (b)}
    \includegraphics[scale=0.6]{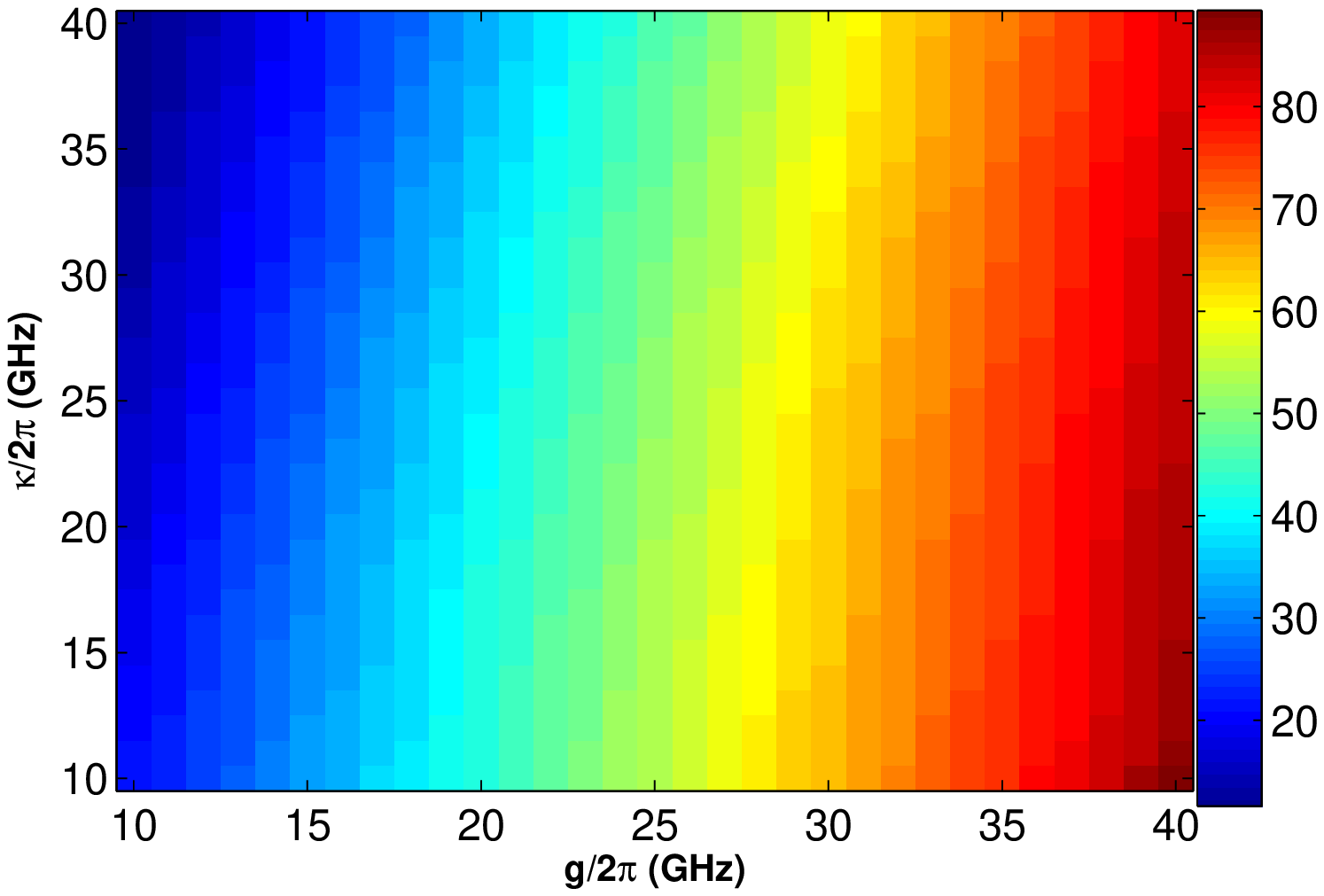}
    %\psgrid
    \rput(-.25,5){\large (b)}
    \includegraphics[scale=0.6]{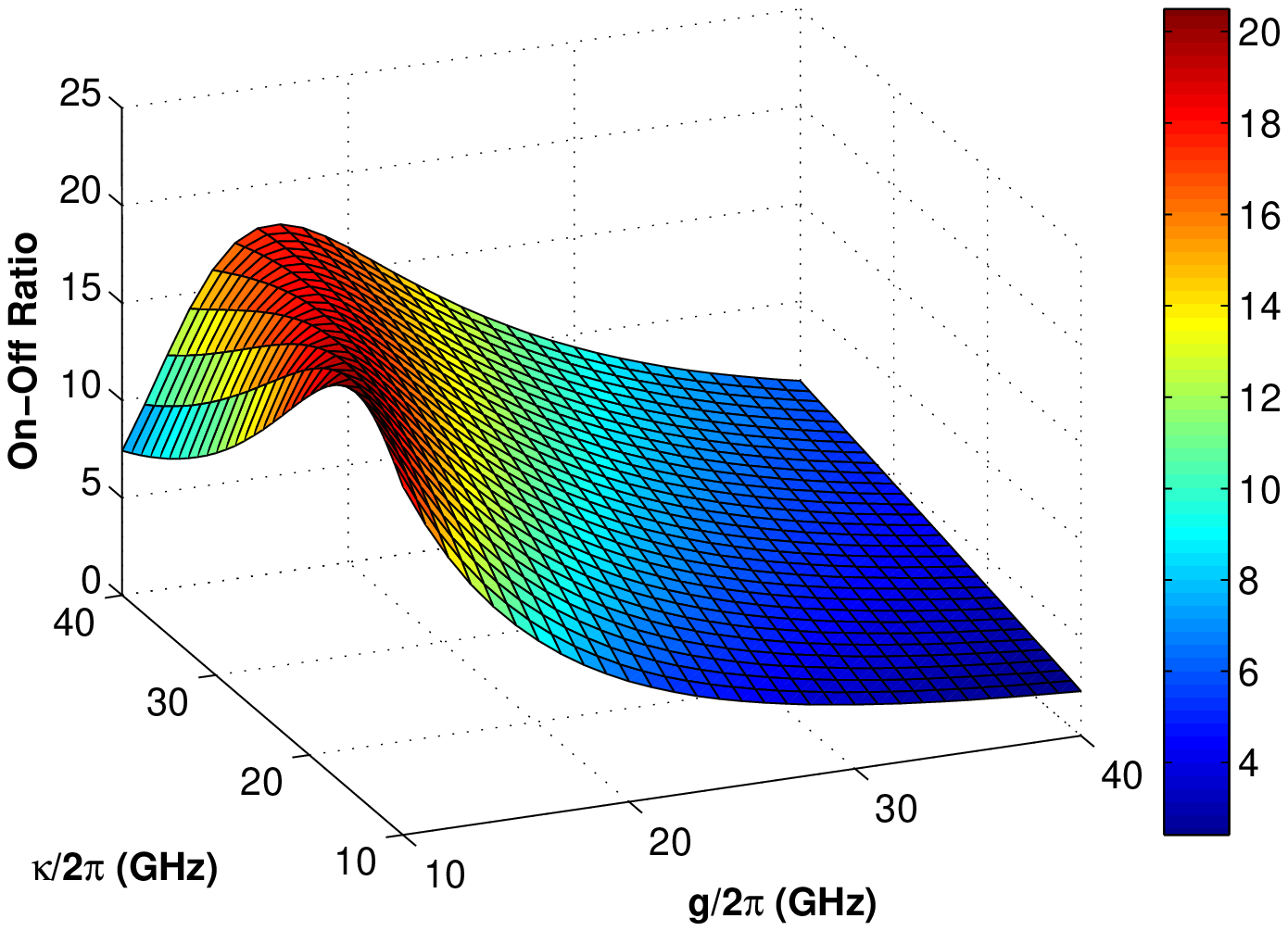}
\caption{ (a) Cut-off frequency of the modulator as a function of
$g$ and $\kappa$; (b) On-off ratio of the modulator at a
modulating frequency of $\omega_e/2\pi = 5$ GHz as a function of
$g$ and $\kappa$. For both plots $\Delta\omega_0/2\pi = 5$ GHz;
$\gamma_d/2\pi = 0.1$ GHz; $\Omega = 1$ GHz. }
\label{cut_off_freq_mbe}
\end{figure}

Fig. \ref{fig:freq_response_mb} shows the on-off ratio of the
output signal as a function of frequency of modulating signal for
different $\kappa$ and $g$. We observe that the modulator behaves
like a second order low pass filter with a roll-off of $-20$
dB/decade. Fig. \ref{cut_off_freq_mbe} shows the cut-off frequency
and the on-off ratio at low frequency as a function of different
$g$ and $\kappa$.

The cut-off frequency of the filter increases with the coupling
strength $g$. Similarly, reduction in $\kappa$ increases the
cut-off frequency. However when $\kappa < g$, the change in
cut-off frequency is not significant and g plays the dominant
role. At the same time,  the on-off ratio decreases with $g$,
which occurs because we kept the maximum detuning
$\Delta\omega_0/2\pi$ fixed at $5$ GHz. By increasing
$\Delta\omega_0/2\pi$ the on-off ratio can be increased for higher
values of $g$.

\section{Nonlinear Distortion of Signal}
\begin{figure}
\centering
    %\includegraphics[scale=1]{fig:freq_response.pdf}
    %\includegraphics[scale=0.7]{fig_outp_sig.eps}
    %\includegraphics[scale=0.7]{fig_outp_sig_high.eps}
    %\psgrid
            \includegraphics[scale=0.6]{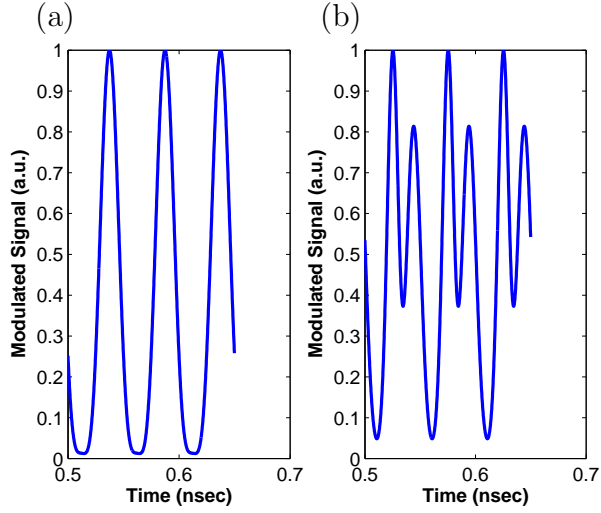}
            \rput(-8,6.6){\large (a)}
            \rput(-4.2,6.6){\large (b)}
\caption{ Normalized output signal for two different  maximal QD
detunings $\Delta\omega_0$. (a) $\Delta\omega_0/2\pi = 2$ GHz. (b)
$\Delta\omega_0/2\pi = 40$ GHz. For both simulations the frequency
of the electrical signal is $\omega_e/2\pi = 20$ GHz, $\kappa/2\pi
= g/2\pi = 20$ GHz and $\Omega = 1$ GHz.} \label{fig:outp_sig}
\end{figure}

\begin{figure}
\centering
    \includegraphics[scale=0.6]{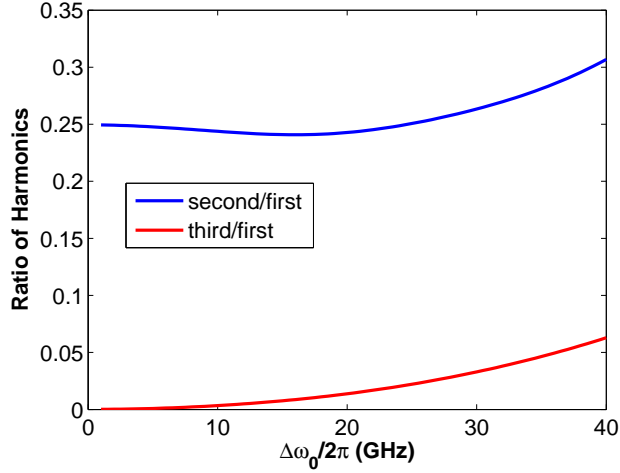}
\caption{Ratio of the second and third harmonic to the first
harmonic of the modulated signal, as a function of the maximum
frequency shift of the QD. The first harmonic is proportional to
the actual signal. For the simulation we assumed that the
modulation is working in the pass-band ($\omega_e/2\pi = 20$ GHz);
$\kappa/2\pi = g/2\pi = 20$ GHz and the dephasing rate
$\gamma_d/2\pi = 0$. } \label{fig:nonlin_distort}
\end{figure}

Another important property of good modulators is that the
modulated optical output should closely resemble the  shape of the
modulating electrical signal. If the modulator does not operate in
the linear regime, the output signal contains spurious higher
harmonics. Fig. \ref{fig:outp_sig} shows the optical output signal
for two different values of maximal QD detuning $\Delta\omega_0$.
For small values of the electrical input (i.e., for small
$\Delta\omega_0$), the output follows exactly the variation of the
QD resonance frequency,  as shown in Fig. \ref{fig:outp_sig}a.
However, distortions appear for higher amplitudes of
$\Delta\omega_0$ (Fig. \ref{fig:outp_sig}b). Here we observe
ripples at a frequency different from the modulating frequency
$\omega_e$. These ripples do not arise only because of the finite
band-width of the system, but are also due to the nonlinearity of
the system with respect to $\Delta\omega_0$. More detailed
analysis of these ripples are described below, when we analyze the
step response of the modulator.
The value of $\Delta\omega_0$ also affects the visibility of the
modulated signal. The on-off ratio (i.e., the desired output) is
proportional to the first harmonic of the modulated output signal.
Fig. \ref{fig:nonlin_distort} shows the ratio of second and third
harmonics to the first harmonic of the output signal as a function
of $\Delta\omega_0$. As expected, the higher order harmonics
increase with increase in QD detuning $\Delta\omega_0$.

\section{Step Response}
\begin{figure}
\centering
    \includegraphics[scale=0.6]{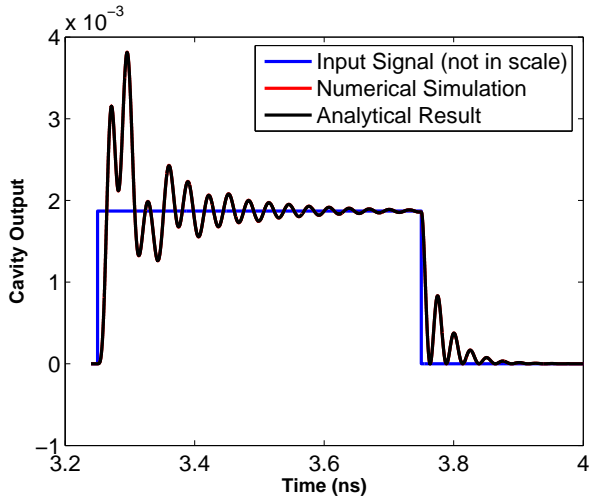}
\caption{ Step response of the single QD electro-optic modulator.
The parameters used for the simulation are: $\kappa/2\pi = 5$ GHz;
$g/2\pi = 20$ GHz; $\Delta\Omega_0 = 20$ GHz; $\Omega =1$ GHz.}
\label{step_response}
\end{figure}

The step response gives information about both the bandwidth of
the system and the nonlinear distortion. To analytically calculate
the step response we consider two situations: ($1$) when the
electrical signal goes from on to off (that is the QD detuning
goes from $\Delta\omega_0 \longrightarrow 0$); and ($2$) when the
electrical signal goes from off to on (that is the QD detuning
goes from $0 \longrightarrow \Delta\omega_0$). For low excitation
and small dephasing rate, the average value of cavity output
$\kappa|\langle a(t)^\dag a(t)\rangle|$ is approximately given by
$\kappa|\langle a(t)\rangle|^2 $, where $\langle a(t)\rangle$ is
\begin{equation}
\langle a (t) \rangle_{\Delta\omega_0 \rightarrow 0} = r(0)
e^{-\alpha(0) t}cos(\beta(0) t +\phi(0)) + SS(0)
\end{equation}
and
\begin{equation}
\langle a (t) \rangle_{0 \rightarrow \Delta\omega_0} =
r(\Delta\omega_0) e^{-\alpha(\Delta\omega_0)
t}cos(\beta(\Delta\omega_0) t +\phi(\Delta\omega_0)) +
SS(\Delta\omega_0)
\end{equation}
where the switching happens at $t=0$ and
\begin{eqnarray*}
  \alpha(\omega) &=& \frac{\kappa+\gamma+i\omega}{2} \\
  \beta(\omega) &=& \frac{\sqrt{4g^2-(\kappa-\gamma-i\omega)^2}}{2} \\
  SS(\omega) &=& \frac{i\Omega(\gamma+i\omega)}{g^2+\kappa(\gamma+i\omega)} \\
  \phi(\omega) &=& tan^{-1}\left(\frac{\alpha(\omega)}{\beta(\omega)}\right)\\
  r(0) &=& \frac{SS(\Delta\omega_0)-SS(0)}{cos(\phi(0))}\\
  r(\Delta\omega_0) &=& \frac{SS(0)-SS(\Delta\omega_0)}{cos(\phi(\Delta\omega_0))}
\end{eqnarray*}
Fig. \ref{step_response} shows the step response of the modulator.
The numerical simulation of MBEs and the analytical expression for
the step response show excellent agreement.
We see the oscillations in step response because the system needs
finite time to relax. Due to non-linearity of the system, this
relaxation time depends on $\Delta\omega_0$. For this reason in
Fig. \ref{fig:outp_sig} we observe ripples only for high values of
$\Delta\omega_0$.
\section{Switching Power}
For most of the modern optoelectronic devices, the energy required
per switching operation is one of the most important figures of
merit \cite{article:miller08}. For the type of devices presented
in this paper, the fundamental limit for the control energy per
switching operation is given by the energy density of the electric
field required to detune the QD inside the active volume. To
estimate the control energy we use the experimental parameters as
presented in \cite{article:Andrei09}. When the electrode is placed
laterally at a distance of $1\mu m$, the active region of the
cavity/QD system has a volume of $V \sim 1\mu m \times 1\mu m
\times 200nm$. Since the electric field to tune the quantum dot is
on the order of $F \sim 5 \times 10^{4}$ V/cm, the energy per
switching operation is on the order of $E=\epsilon_{0}
\epsilon_{r} F^{2}V/2 \sim 0.5$ fJ ($\epsilon_{r} \sim 13$). This
translates into an operating power of $\sim 5 \mu $ W at $10$ GHz.
The same order of magnitude estimation is obtained by modeling the
device as a parallel plate capacitor with width $w \sim 1\mu m$,
thickness $t \sim 200nm$ and spacing $L \sim 1\mu m$ and taking
into account fringing effects. Since the energy consumption has a
quadratic dependence on the applied voltage, the operating power
can be lowered significantly by bringing the electrode closer to
the quantum dot. In case the active volume could be reduced to the
size of the quantum dot itself ($\sim 25 \times 25 \times 25$
$nm^3$), the switching energy can be lowered below $0.1$ aJ. These
energy scales are of the same order of magnitude as all other
optical switching devices operating at single photon level
\cite{article:eng07, article:fushman08}, and are orders of
magnitude lower than the current state of the art electro-optic
modulators
\cite{2008.Liu.NatPhot.50fJModulator,2008.Xu.HP.MicronScaleSiMicroring}.

\section{Effect of Dephasing of QD}
\begin{figure}
\centering
    \rput(-.2,4.7){\large (a)}
    \rput(6.4,4.7){\large (b)}
    \includegraphics[scale=0.4]{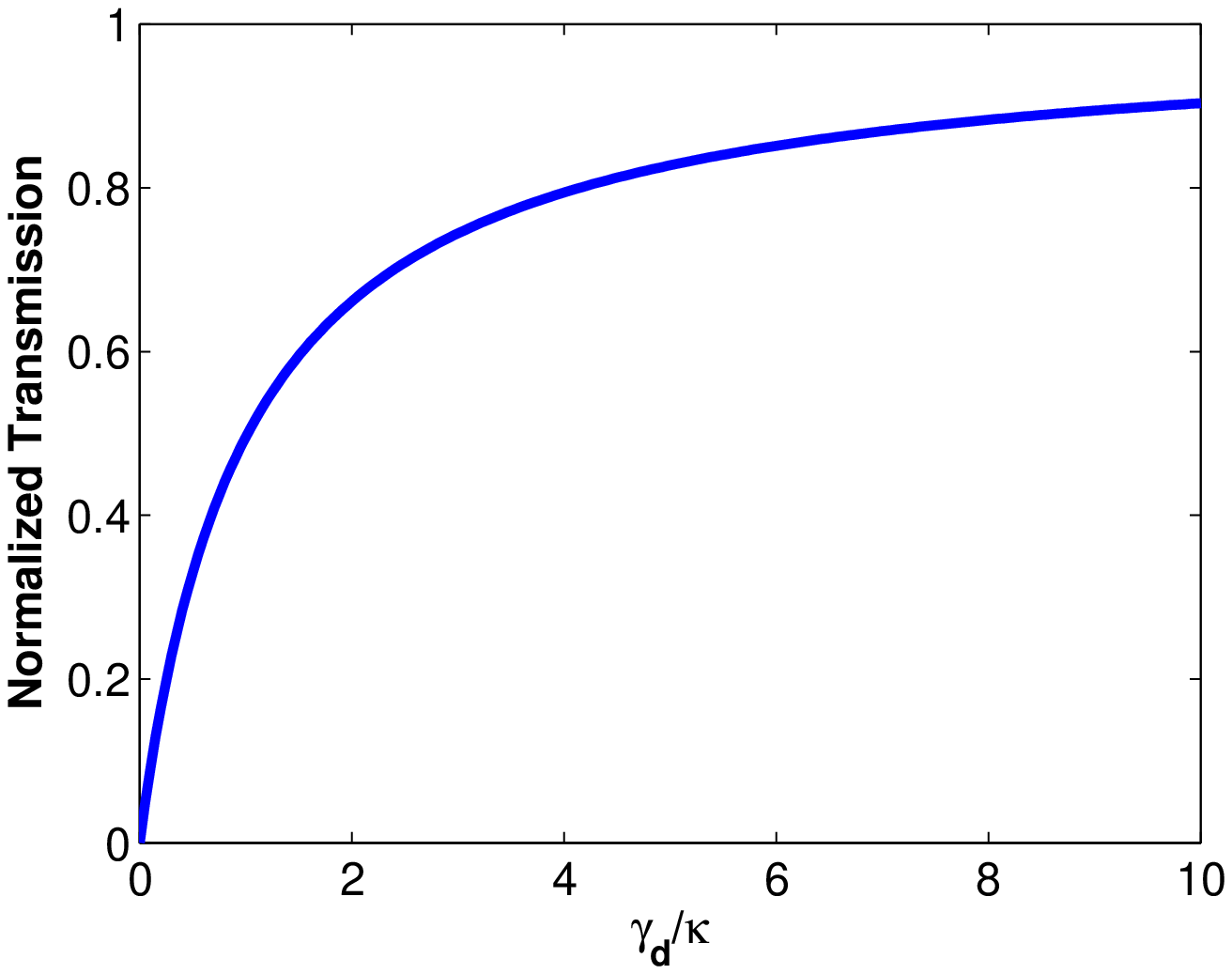} \includegraphics[scale=0.4]{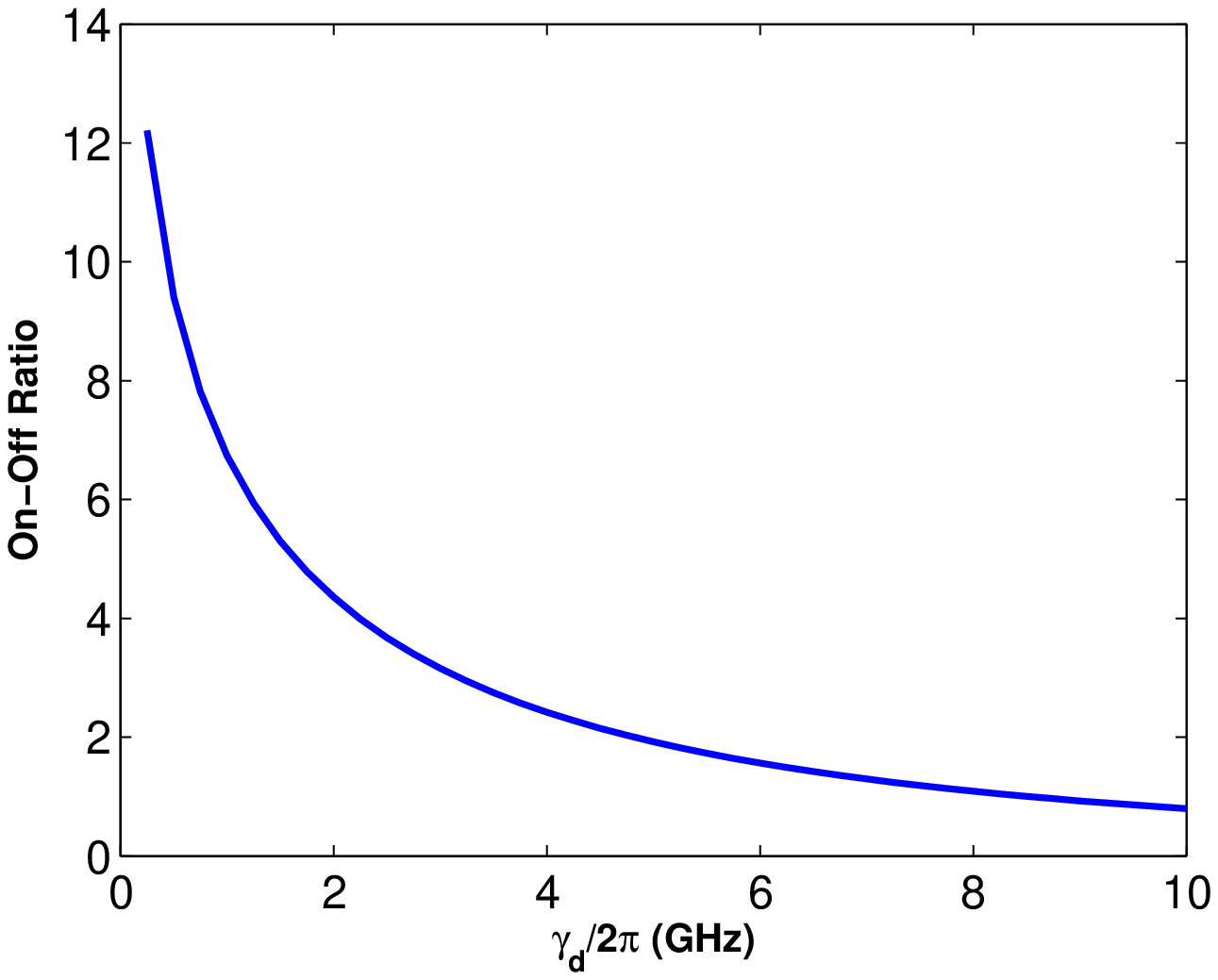}
\caption{(a) Normalized steady state transmission of the laser
resonant with the dot (i.e., $\Delta\omega_a=0$) with pure
dephasing rate $\gamma_d/2\pi$. Parameters used are: $g/2\pi =
\kappa/2\pi = 20$ GHz; $\gamma = \kappa/80$; (b) On-off ratio of
the modulated signal as a function of the dephasing rate, for
$\kappa/2\pi = g/2\pi = 20$ GHz. The modulation frequency
$\omega_e/2\pi = 5$ GHz and the amplitude of the change in
resonance frequency $\Delta\omega_0/2\pi = 10$ GHz.}
\label{fig:effect_dephasing}
\end{figure}

One of the major problems in CQED with a QD is the dephasing of
the QD, caused by interaction of the QD with nearby nuclei and
phonons. The dip in transmission through the cavity in presence of
a coupled QD is caused by the destructive interference of the
incoming light and the light absorbed and re-radiated by the QD
\cite{article:edo_dit}. Due to the dephasing of the QD, the light
scattered from the QD is not exactly out-of-phase with the
incoming light. This affects the destructive interference thus
causing the cavity to be not fully reflective. Fig.
\ref{fig:effect_dephasing}a shows the steady state transmission
through a coupled cavity/QD system (normalized by the transmission
through an empty cavity) of a laser resonant with the cavity and
the QD, as a function of QD dephasing rate. Hence, the performance
of the modulator also suffers because of dephasing. Fig.
\ref{fig:effect_dephasing}b shows the on-off ratio as a function
of the dephasing rate $\gamma_d/2\pi$ of the QD. As expected the
on-off ratio falls off with increased dephasing rate.
\section{Conclusion}
We have performed the detailed analysis of the performance of an
electro-optic modulator where the modulation is provided by a
single QD strongly coupled to a cavity. We have shown that a
modulation speed of $40$ GHz can be achieved for realistic system
parameters ($\kappa/2\pi = g/2\pi = 20$ GHz) with a control energy
of the order of $0.5$ fJ. This type of fast and low control energy
modulator may constitute an essential building block of future
nano-photonic networks.
\section{Acknowledgement}
The authors gratefully acknowledge financial support provided by
the National Science Foundation, and Office of Naval Research.
A.M. was supported by the Stanford Graduate Fellowship (Texas
Instruments fellowship).

\end{document}